# The Multiphoton Boson Sampling Machine Doesn't Beat Early Classical Computers for Five-boson Sampling

Shenghui Su [1, 3, 4] and Jianhua Zheng [2]

[1] College of Computers, Nanjing Univ. of Aeronautics & Astronautics, Nanjing 211106, PRC
[2] School of Network Securities, Information Engineering University, Zhengzhou 450001, PRC
[3] Pub-sec Innovation Center, Nanjing Univ. of Science and Technology, Nanjing 210094, PRC
[4] Laboratory of Computational Complexity, BFID Corporation, Beijing 100098, PRC
{Corresponding email: reesse@126.com / Received: 08 Feb 2018 & Revised: 03 Jan 2019}

**Abstract**: A new algorithm which is called Store-zechin, and utilizes stored data repetitively for calculating the permanent of an $n \times n$ matrix is proposed. The analysis manifests that the numbers of multiplications and additions taken by the new algorithm are respectively far smaller than those taken by the famous Ryser algorithm. Especially, for a 5-boson sampling task, the running time of the Store-zechin algorithm computing the correspondent permanent on ENIAC as well as TRADIC is lower than that of the sampling operation on a multiphoton boson sampling machine (shortly MPBSM), and thus MPBSM *does not* beat the early classical computers (despite of this, it is possible that when $n$ gets large enough, a quantum boson sampling machine will beat a classical computer). On a computer, people can design an algorithm that exchanges space for time while on MPBSM, people can not do so, which is the greatest difference between a universal computer and MPBSM. This difference is right the reason why MPBSM may not be called a (photonic) quantum computer.

## 1 Introduction

Around April 2017, a group of scholars constructed successfully a multiphoton boson sampling machine (shortly MPBSM), demonstrated a 3-photon boson sampling rate of 4.96 kHz, and achieved 4- and 5-photon boson samplings using a particular single-photon state [1][2]. Moreover, they claimed that MPBSM could beat early classical computers, and be provably faster respectively than ENIAC and TRADIC, the first electronic computer and the first transistorized computer, for the boson sampling tasks [1][2].

MPBSM samples the output distributions of 3, 4, and 5 bosons separately in practice, and thus it is essentially a 5-photon boson sampling machine. The experimental and analytic data of Reference [1] and [2] show that MPBSM beats both ENIAC and TRADIC for 3- and 4-photon boson sampling tasks, and beats only ENIAC for a 5-photon boson sampling task. If indeed a fact, it is logical that MPBSM beats the early classical computers.

However, unfortunately, MPBSM beats neither ENIAC nor TRADIC factually for a 5-photon boson sampling task. Summarily, MPBSM does not beat the early classical computers. In this paper, we will show that according to a new algorithm idea, MPBSM does not beat ENIAC for a 5-photon boson sampling task.

## 2 Boson Sampling Machine and Permanent of a Square Matrix

A photonic quantum boson sampling machine is a type of quantum simulator which has no memory, requires neither nonlinearities nor qubit entanglement, and performs nothing other than a special task — boson sampling for example [3][4]. Unlike a universal quantum computer, a boson sampling machine (shortly BSM) consists of merely a single-photon source, a multiphoton interferometer, and a set of detectors. The first 3- and 4-photon BSM of the world are created in 2012 separately by two teams [4][5], and used to illustrate that a quantum simulator could solve a specific problem more efficiently than a classical computer, and consequently a universal quantum computer is worth researching and developing.

A BSM samples the output distribution of $n$ indistinguishable bosons scattered by a unitary transformation $U$, an integrated photonic circuit. That is to say, to repeatedly inject $n$ photons in an input state, detect all outputs, collect all $n$-fold coincident events, and make the statistic of all valid outputs by the values or states [4][5]. Mathematically, the probability amplitude of an output result is directly proportional to the permanent of a corresponding $n \times n$ submatrix of $U$ [4][5]. Calculating the permanent of a nonnegative integral (even 0, 1) square matrix is proved to be one #P-complete problem which belongs to a class of problems believed to be even more difficult to compute than NP in computational complexity theory [6][7], and therefore calculating the permanent is at least exponentially intractable up to the present.

## 3  Definition and Calculation of a Permanent

There is an apparent similarity between the definition of a permanent and a determinant.

### 3.1  Definition and Properties of a Permanent

The permanent of an $n \times n$ matrix $A = (a_{i,j})$ is defined as

$$\text{per}(A) = \sum_{\sigma \in S_n} \prod_{i=1}^{n} a_{i,\sigma(i)}, \tag{1}$$

where $S_n$ is the symmetric group over the set $\{1, 2, \ldots, n\}$, and $\sigma$ is an element of $S_n$, namely a permutation of the numbers $1, 2, \ldots, n$ [8][9]. This formula is similar to the corresponding formula for the determinant except the sign of each permutation $\sigma$ influencing the $\pm$ polarity of a product term.

Given a square matrix $A = (a_{i,j})$ of order $n$, its permanent has the following properties frequently used [8][9]:

① per($A$) is invariant under a permutation of two arbitrary rows or columns of $A$,

② per($A$) is invariant under the transposition of $A$, that is, per($A$) = per($A^\top$),

③ per($A$) will be changed to $s \cdot$per($A$) when any row or column of $A$ is multiplied by a scalar $s$,

④ per($A$) may be developed along a row or column of $A$, which is similar to the Laplace's development of a determinant, but ignores all signs.

For example, developing along the first column,

$$\text{per}\begin{pmatrix} 7 & 0 & 1 & 2 \\ 5 & 3 & 4 & 5 \\ 3 & 5 & 6 & 7 \\ 1 & 7 & 8 & 9 \end{pmatrix} = 7\cdot\text{per}\begin{pmatrix} 3 & 4 & 5 \\ 5 & 6 & 7 \\ 7 & 8 & 9 \end{pmatrix} + 5\cdot\text{per}\begin{pmatrix} 0 & 1 & 2 \\ 5 & 6 & 7 \\ 7 & 8 & 9 \end{pmatrix} + 3\cdot\text{per}\begin{pmatrix} 0 & 1 & 2 \\ 3 & 4 & 5 \\ 7 & 8 & 9 \end{pmatrix} + 1\cdot\text{per}\begin{pmatrix} 0 & 1 & 2 \\ 3 & 4 & 5 \\ 5 & 6 & 7 \end{pmatrix}.$$

Further, the permanents of matrices of order 3 may be developed to those of matrices of order 2.

### 3.2  Naive Algorithm for Calculating Permanents

It is well known that calculating the permanent of a square matrix is more difficult than calculating the determinant of a square matrix.

According to Formula (1), one can directly acquire a computer algorithm called Naive which will take $n!(n - 1)$ multiplications and $(n! - 1)$ additions because the order of $S_n$ is $n!$, and each product involves $n$ elements at a different row and column of the matrix.

## 3.3 Ryser Algorithm for Calculating Permanents

The most efficient method of calculating a permanent so far was proposed by Ryser in 1963 [10]. The Ryser method resorts to an inclusion-exclusion manner by which per($A$) can be expressed as

$$\text{per}(A) = \sum_{k=0}^{n-1} (-1)^k T_k, \tag{2}$$

where $T_k$ is the sum of the values of $P(A_k)$ over all possible $A_k$, $A_k$ is a matrix obtained from $A$ with columns $k$ removed, and $P(A_k)$ is the product of the row-sums of $A_k$.

It is not difficult to see that in terms of the matrix entries, Formula (2) may be rewritten as

$$\text{per}(A) = \sum_{\varepsilon_1 \ldots \varepsilon_n \in \{0,1\}^n} (-1)^{n-\varepsilon_1-\ldots-\varepsilon_n} \prod_{i=1}^{n} \sum_{j=1}^{n} \varepsilon_j a_{ij}, \tag{3}$$

which seemingly has some extra operations.

Besides, Formula (2) may also be translated to

$$\text{per}(A) = (-1)^n \sum_{S \subseteq \{1,\ldots,n\}} (-1)^{|S|} \prod_{i=1}^{n} \sum_{j \in S} a_{ij}, \tag{4}$$

where $|S|$ is the element number of a subset $S$, and the selection of $S$ should be in such an ascending order of $|S|$ (= 1, 2, …, $n$) that a previous $\sum_{j \in S} a_{ij}$ may be exploited — ($a_{11} + a_{12}$) may be exploited by ($a_{11} + a_{12} + a_{13}$) for example.

According to Formula (2), (3) or (4), one can acquire a computer algorithm called Ryser for calculating a permanent. Since the number of possible subsets (excluding the empty) of $\{1, \ldots, n\}$ is $2^n - 1$, $|S|$'s are in the ascending order, and a previous $\sum_{j \in S} a_{ij}$ with $|S| = x$ may be reused by a current $\sum_{j \in S} a_{ij}$ with $|S| = x+1$, we can exactly evaluate the numbers of multiplications and additions required by the Ryser algorithm respectively as $(2^n - 1)(n - 1)$ and $(2^n - n)(n + 1) - 2$.

Notice that in Reference [1] and [2], the number of additions is evaluated as $(2^n - 2)(n + 1)$ which is bigger than $(2^n - n)(n + 1) - 2$ although $\lim_{n \to \infty}((2^n - n)(n + 1) - 2) / ((2^n - 2)(n + 1)) = 1$.

## 4 Store-zechin Algorithm for Calculating Permanents

A new algorithm called Store-zechin which has been seemingly ignored by mathematicians for a long time is put forward. It employs the storage of a computer sufficiently, and utilizes the stored data (like zechins) repetitively, by which per($A$) is developed along the last row up to the permanents of matrices of order 2.

For convenience, we denote by $A_{u;\,v}$ a square matrix $A$ with rows $u$ and columns $v$ removed.

### 4.1 Case of a 3 × 3 Matrix

Let a 3 × 3 matrix

$$A = \begin{pmatrix} a_{11} & a_{12} & a_{13} \\ a_{21} & a_{22} & a_{23} \\ a_{31} & a_{32} & a_{33} \end{pmatrix},$$

and then

$$A_{3;\,1} = \begin{pmatrix} a_{12} & a_{13} \\ a_{22} & a_{23} \end{pmatrix}, \quad A_{3;\,2} = \begin{pmatrix} a_{11} & a_{13} \\ a_{21} & a_{23} \end{pmatrix}, \quad A_{3;\,3} = \begin{pmatrix} a_{11} & a_{12} \\ a_{21} & a_{22} \end{pmatrix}.$$

Therefore, we have

$$\text{per}(A) = a_{31} \text{per}(A_{3;\,1}) + a_{32} \text{per}(A_{3;\,2}) + a_{33} \text{per}(A_{3;\,3})$$
$$= a_{31}(a_{12} a_{23} + a_{13} a_{22}) + a_{32}(a_{11} a_{23} + a_{13} a_{21}) + a_{33}(a_{11} a_{22} + a_{12} a_{21}).$$

The above development may be directly transformed into a specific algorithm of Store-zechin for per($A$ of order 3). The arithmetic steps of the algorithm can be analyzed.

| At Layer 1 | Number of Multiplications | Number of Additions |
|---|---|---|
| Term 1 | 3 | 1 |
| Term 2 | 3 | 1 |
| Term 3 | 3 | 1 |
| Σ | 0 | 2 |
| Total | 9 | 5 |

Table 1: Numbers of ×'s and +'s Taken by the Algorithm for Per($A$ of Order 3)

It should be noted that in the development of permanent of a 3 × 3 matrix, there is no repeating term, but some common factors are extracted.

### 4.2  Case of a 4 × 4 Matrix

Let a 4 × 4 matrix

$$A = \begin{pmatrix} a_{11} & a_{12} & a_{13} & a_{14} \\ a_{21} & a_{22} & a_{23} & a_{24} \\ a_{31} & a_{32} & a_{33} & a_{34} \\ a_{41} & a_{42} & a_{43} & a_{44} \end{pmatrix},$$

and then

$$A_{4;1} = \begin{pmatrix} a_{12} & a_{13} & a_{14} \\ a_{22} & a_{23} & a_{24} \\ a_{32} & a_{33} & a_{34} \end{pmatrix} \quad A_{4;2} = \begin{pmatrix} a_{11} & a_{13} & a_{14} \\ a_{21} & a_{23} & a_{24} \\ a_{31} & a_{33} & a_{34} \end{pmatrix} \quad A_{4;3} = \begin{pmatrix} a_{11} & a_{12} & a_{14} \\ a_{21} & a_{22} & a_{24} \\ a_{31} & a_{32} & a_{34} \end{pmatrix}$$

$$A_{4;4} = \begin{pmatrix} a_{11} & a_{12} & a_{13} \\ a_{21} & a_{22} & a_{23} \\ a_{31} & a_{32} & a_{33} \end{pmatrix}$$

Therefore, we have

$\text{per}(A) = a_{41} \text{per}(A_{4;1}) + a_{42} \text{per}(A_{4;2}) + a_{43} \text{per}(A_{4;3}) + a_{44} \text{per}(A_{4;4})$
$= a_{41}(a_{32} \text{per}(A_{3,4;1,2}) + a_{33} \text{per}(A_{3,4;1,3}) + a_{34} \text{per}(A_{3,4;1,4})) +$
$\quad a_{42}(a_{31} \text{per}(A_{3,4;1,2}) + a_{33} \text{per}(A_{3,4;2,3}) + a_{34} \text{per}(A_{3,4;2,4})) +$
$\quad a_{43}(a_{31} \text{per}(A_{3,4;1,3}) + a_{32} \text{per}(A_{3,4;2,3}) + a_{34} \text{per}(A_{3,4;3,4})) +$
$\quad a_{44}(a_{31} \text{per}(A_{3,4;1,4}) + a_{32} \text{per}(A_{3,4;2,4}) + a_{33} \text{per}(A_{3,4;3,4})).$

The above development may be transformed into a specific algorithm of Store-zechin for per($A$ of order 4). Also, the arithmetic steps of the algorithm can be analyzed.

| At Layer 1 | Number of Multiplications | Number of Additions |
|---|---|---|
| Term 1 | 10 | 5 |
| Term 2 | 8 | 4 |
| Term 3 | 6 | 3 |
| Term 4 | 4 | 2 |
| Σ | 0 | 3 |
| Total | 28 | 17 |

Table 2: Numbers of ×'s and +'s Taken by the Algorithm for Per($A$ of Order 4)

It should be noted that in the development of permanent of a $4 \times 4$ matrix, at Layer 2 there are 6 two-fold repeating terms, and each repeating term is calculated one time only.

### 4.3  Case of a $5 \times 5$ Matrix

Let a $5 \times 5$ matrix

$$A = \begin{pmatrix} a_{11} & a_{12} & a_{13} & a_{14} & a_{15} \\ a_{21} & a_{22} & a_{23} & a_{24} & a_{25} \\ a_{31} & a_{32} & a_{33} & a_{34} & a_{35} \\ a_{41} & a_{42} & a_{43} & a_{44} & a_{45} \\ a_{51} & a_{52} & a_{53} & a_{44} & a_{55} \end{pmatrix}$$

and then

$$A_{5;1} = \begin{pmatrix} a_{12} & a_{13} & a_{14} & a_{15} \\ a_{22} & a_{23} & a_{24} & a_{25} \\ a_{32} & a_{33} & a_{34} & a_{35} \\ a_{42} & a_{43} & a_{44} & a_{45} \end{pmatrix} \quad A_{5;2} = \begin{pmatrix} a_{11} & a_{13} & a_{14} & a_{15} \\ a_{21} & a_{23} & a_{24} & a_{25} \\ a_{31} & a_{33} & a_{34} & a_{35} \\ a_{41} & a_{43} & a_{44} & a_{45} \end{pmatrix} \quad A_{5;3} = \begin{pmatrix} a_{11} & a_{12} & a_{14} & a_{15} \\ a_{21} & a_{22} & a_{24} & a_{25} \\ a_{31} & a_{32} & a_{34} & a_{35} \\ a_{41} & a_{42} & a_{43} & a_{45} \end{pmatrix}$$

$$A_{5;4} = \begin{pmatrix} a_{11} & a_{12} & a_{13} & a_{15} \\ a_{21} & a_{22} & a_{23} & a_{25} \\ a_{31} & a_{32} & a_{33} & a_{35} \\ a_{41} & a_{42} & a_{43} & a_{45} \end{pmatrix} \quad A_{5;5} = \begin{pmatrix} a_{11} & a_{12} & a_{13} & a_{14} \\ a_{21} & a_{22} & a_{23} & a_{24} \\ a_{31} & a_{32} & a_{33} & a_{34} \\ a_{41} & a_{42} & a_{43} & a_{44} \end{pmatrix}$$

Therefore, we have

$\text{per}(A) = a_{51} \text{per}(A_{5;1}) + a_{52} \text{per}(A_{5;2}) + a_{53} \text{per}(A_{5;3}) + a_{54} \text{per}(A_{5;4}) + a_{55} \text{per}(A_{5;5})$
$= a_{51}(a_{42} \text{per}(A_{4,5;1,2}) + a_{43} \text{per}(A_{4,5;1,3}) + a_{44} \text{per}(A_{4,5;1,4}) + a_{45} \text{per}(A_{4,5;1,5})) +$
$\quad a_{52}(a_{41} \text{per}(A_{4,5;1,2}) + a_{43} \text{per}(A_{4,5;2,3}) + a_{44} \text{per}(A_{4,5;2,4}) + a_{45} \text{per}(A_{4,5;2,5})) +$
$\quad a_{53}(a_{41} \text{per}(A_{4,5;1,3}) + a_{42} \text{per}(A_{4,5;2,3}) + a_{44} \text{per}(A_{4,5;3,4}) + a_{45} \text{per}(A_{4,5;3,5})) +$
$\quad a_{54}(a_{41} \text{per}(A_{4,5;1,4}) + a_{42} \text{per}(A_{4,5;2,4}) + a_{43} \text{per}(A_{4,5;3,4}) + a_{45} \text{per}(A_{4,5;4,5})) +$
$\quad a_{55}(a_{41} \text{per}(A_{4,5;1,5}) + a_{42} \text{per}(A_{4,5;2,5}) + a_{43} \text{per}(A_{4,5;3,5}) + a_{44} \text{per}(A_{4,5;4,5}))$
$= a_{51}(a_{42}(a_{33} \text{per}(A_{3,4,5;1,2,3}) + a_{34} \text{per}(A_{3,4,5;1,2,4}) + a_{35} \text{per}(A_{3,4,5;1,2,5})) +$
$\quad\quad a_{43}(a_{32} \text{per}(A_{3,4,5;1,2,3}) + a_{34} \text{per}(A_{3,4,5;1,3,4}) + a_{35} \text{per}(A_{3,4,5;1,3,5})) +$
$\quad\quad a_{44}(a_{32} \text{per}(A_{3,4,5;1,2,4}) + a_{33} \text{per}(A_{3,4,5;1,3,4}) + a_{35} \text{per}(A_{3,4,5;1,4,5})) +$
$\quad\quad a_{45}(a_{32} \text{per}(A_{3,4,5;1,2,5}) + a_{33} \text{per}(A_{3,4,5;1,3,5}) + a_{34} \text{per}(A_{3,4,5;1,4,5}))) +$
$\quad a_{52}(a_{41}(a_{33} \text{per}(A_{3,4,5;1,2,3}) + a_{34} \text{per}(A_{3,4,5;1,2,4}) + a_{35} \text{per}(A_{3,4,5;1,2,5})) +$
$\quad\quad a_{43}(a_{31} \text{per}(A_{3,4,5;1,2,3}) + a_{34} \text{per}(A_{3,4,5;2,3,4}) + a_{35} \text{per}(A_{3,4,5;2,3,5})) +$
$\quad\quad a_{44}(a_{31} \text{per}(A_{3,4,5;1,2,4}) + a_{33} \text{per}(A_{3,4,5;2,3,4}) + a_{35} \text{per}(A_{3,4,5;2,4,5})) +$
$\quad\quad a_{45}(a_{31} \text{per}(A_{3,4,5;1,2,5}) + a_{33} \text{per}(A_{3,4,5;2,3,5}) + a_{34} \text{per}(A_{3,4,5;2,4,5}))) +$
$\quad a_{53}(a_{41}(a_{32} \text{per}(A_{3,4,5;1,2,3}) + a_{34} \text{per}(A_{3,4,5;1,3,4}) + a_{35} \text{per}(A_{3,4,5;1,3,5})) +$
$\quad\quad a_{42}(a_{31} \text{per}(A_{3,4,5;1,2,3}) + a_{34} \text{per}(A_{3,4,5;2,3,4}) + a_{35} \text{per}(A_{3,4,5;2,3,5})) +$
$\quad\quad a_{44}(a_{31} \text{per}(A_{3,4,5;1,3,4}) + a_{32} \text{per}(A_{3,4,5;2,3,4}) + a_{35} \text{per}(A_{3,4,5;3,4,5})) +$
$\quad\quad a_{45}(a_{31} \text{per}(A_{3,4,5;1,3,5}) + a_{32} \text{per}(A_{3,4,5;2,3,5}) + a_{34} \text{per}(A_{3,4,5;3,4,5}))) +$
$\quad a_{54}(a_{41}(a_{32} \text{per}(A_{3,4,5;1,2,4}) + a_{33} \text{per}(A_{3,4,5;1,3,4}) + a_{35} \text{per}(A_{3,4,5;1,4,5})) +$
$\quad\quad a_{42}(a_{31} \text{per}(A_{3,4,5;1,2,4}) + a_{33} \text{per}(A_{3,4,5;2,3,4}) + a_{35} \text{per}(A_{3,4,5;2,4,5})) +$
$\quad\quad a_{43}(a_{31} \text{per}(A_{3,4,5;1,3,4}) + a_{32} \text{per}(A_{3,4,5;2,3,4}) + a_{35} \text{per}(A_{3,4,5;3,4,5})) +$
$\quad\quad a_{45}(a_{31} \text{per}(A_{3,4,5;1,4,5}) + a_{32} \text{per}(A_{3,4,5;2,4,5}) + a_{33} \text{per}(A_{3,4,5;3,4,5}))) +$
$\quad a_{55}(a_{41}(a_{32} \text{per}(A_{3,4,5;1,2,5}) + a_{33} \text{per}(A_{3,4,5;1,3,5}) + a_{34} \text{per}(A_{3,4,5;1,4,5})) +$
$\quad\quad a_{42}(a_{31} \text{per}(A_{3,4,5;1,2,5}) + a_{33} \text{per}(A_{3,4,5;2,3,5}) + a_{34} \text{per}(A_{3,4,5;2,4,5})) +$

$a_{43}(a_{31} \text{per}(A_{3, 4, 5; 1, 3, 5}) + a_{32} \text{per}(A_{3, 4, 5; 2, 3, 5}) + a_{34} \text{per}(A_{3, 4, 5; 3, 4, 5})) +$

$a_{44}(a_{31} \text{per}(A_{3, 4, 5; 1, 4, 5}) + a_{32} \text{per}(A_{3, 4, 5; 2, 4, 5}) + a_{33} \text{per}(A_{3, 4, 5; 3, 4, 5})))$.

The above development may be transformed into a specific algorithm of Store-zechin for per($A$ of order 5). The arithmetic steps of the algorithm can be analyzed.

| At Layer 1 | Number of Multiplications | Number of Additions |
| --- | --- | --- |
| Term 1 | 29 | 17 |
| Term 2 | 20 | 12 |
| Term 3 | 13 | 8 |
| Term 4 | 8 | 5 |
| Term 5 | 5 | 3 |
| Σ | 0 | 4 |
| Total | 75 | 49 |

Table 3: Numbers of ×'s and +'s Taken by the Algorithm for Per($A$ of Order 5)

It should be noted that in the development of permanent of a 5 × 5 matrix, at Layer 2 there are 10 two-fold repeating terms, and at Layer 3 there are 10 six-fold repeating terms, and each repeating term is only calculated one time.

## 5 Comparison between Classical Running Times and MPBSM Running Times

People usually measure the running time or time complexity of an algorithm by the arithmetic steps, and it is a scale independent of a computer CPU.

### 5.1 Arithmetic Steps of Three Algorithms for Calculating Permanents

Through the above discussion, we know that there are the three algorithms for calculating the permanent of an $n \times n$ matrix. They each can be executed on a classical computer.

It can be seen from Section 3 that the naive algorithm takes $n!(n - 1)$ multiplications and $(n! - 1)$ additions, and the Ryser algorithm takes $(2^n - 1)(n - 1)$ multiplications and $(2^n - n)(n + 1) - 2$ (or $(2^n - 2)(n + 1)$) additions. At present, there is no general formula yet for the arithmetic steps of the Store-zechin algorithm, but specific arithmetic steps can be obtained as $n = 3, 4,$ and 5. Hence, the following list can be drawn.

| Algorithm Name | $n = 3$ | | $n = 4$ | | $n = 5$ | |
| --- | --- | --- | --- | --- | --- | --- |
| | Number of ×'s | Number of +'s | Number of ×'s | Number of +'s | Number of ×'s | Number of +'s |
| Naive | 12 | 5 | 72 | 23 | 480 | 119 |
| Ryser | 14 | 18 ( 24) | 45 | 58 (70) | 124 | 160 (180) |
| Store-zechin | 9 | 5 | 28 | 17 | 75 | 49 |

Table 4: Respective Arithmetic Steps of the Three Algorithms

Notice that the numbers parenthesized on Table 4 are evaluated by the formula $(2^n - 2)(n + 1)$.

Obviously, the Store-zechin algorithm has an advantage over the Naive algorithm and Ryser algorithm.

## 5.2 Running Times of Classical Computers and Running Times of MPBSM

On a concrete computer, the concrete running time of an algorithm can be acquired.

In terms of Reference [11] and [12], the first electronic computer, ENIAC, performs 5000 additions or 357 multiplications per second, and the first transistorized computer, TRADIC, performs 62500 additions or 3333 multiplications per second.

The boson sampling itself is not a #P-complete problem, but relates to (exactly speaking, is directly proportional to) the calculation of permanent of an $n \times n$ matrix. In conformity to Reference [1] and [2], the running time of a boson sampling operation on MPBSM can be obtained, in conformity to Reference [11] and [12], the running time of an algorithm on ENIAC or TRADIC can also be obtained, and therefore a comparison between a classical running time and a MPBSM running time can be made.

Note that when a square matrix contains negative or nonintegral elements, the calculation of its permanent is not a #P-complete problem once more (may resort to the Gurvits's approximation algorithm with a polynomial running time [13]), and under this circumstance, the comparison between a Gurvits's running time and a MPBSM running time is illogical; thus we do not consider the Gurvits's algorithm in the following comparisons.

|  |  | 3-Boson Sampling $\propto$ Per(3×3 Submatrix) | 4-Boson Sampling $\propto$ Per(4×4 Submatrix) | 5-Boson Sampling $\propto$ Per(5×5 Submatrix) |
|---|---|---|---|---|
| Naive | ENIAC | 34.6 | 206.3 | 1368.3 |
|  | TRADIC | 3.7 | 22.0 | 145.9 |
| Ryser | ENIAC | 42.8 (44.0) | 137.7 (140.0) | 379.3 (383.3) |
|  | TRADIC | 4.5 (4.6) | 14.4 (14.6) | 39.8 (40.1) |
| Store-zechin | ENIAC | 26.2 | 81.8 | 219.9 |
|  | TRADIC | 2.8 | 8.7 | 23.3 |
| Quantum | MPBSM | 0.2 | 6.6 | 248.8 |

Table 5: Classical Running Times and MPBSM Running Times (Unit: milliseconds)

Notice that the data parenthesized on Table 5 relate to the formula $(2^n - 2)(n + 1)$, and running times of MPBSM are from Ref. [1] & [2].

We see from Table 5 that when $n = 5$, the running times 219.9 and 23.3 of the Store-zechin algorithm on ENIAC and TRADIC are individually smaller than the running time 248.8 of the 5-photon boson sampling operation on MPBSM. Of course, people should select the fastest algorithm as a benchmarking algorithm. Hence, MPBSM *does not* beat the early classical computers ENIAC and TRADIC with regard to a 5-photon boson sampling task.

## 6 Conclusions

On a computer, people can design an algorithm that exchanges space for time while on a photonic quantum boson sampling machine, people can not do so, which is the greatest difference between a universal computer and a boson sampling machine. This difference is right the reason why a photonic quantum boson sampling machine *may not* be called a (photonic) quantum computer.

In the paper, a new algorithm called Store-zechin is considered, and its characteristic is the sufficient employment of computer storages and the repetitive utilization of stored dada (like zechins). For a 5-photon boson sampling task, the running time of the Store-zechin algorithm calculating the correspondent permanent on ENIAC or TRADIC is lower than that of the boson sampling operation on MPBSM, and therefore, in summary, MPBSM does not beat the early classical computers.

It is thought possible that when $n$ gets sufficiently large — at least 32 for example, a photonic quantum boson sampling machine will beat present classical computers or supercomputers, which is used to demonstrate that the Extended Church-Turing thesis is incorrect. However, it is more convincing that a universal quantum computer will demonstrate that an integer factorization problem or a discrete logarithm problem can be solved in polynomial time.

## Acknowledgment

The authors would like to thank the Acad. Jiren Cai, Acad. Zhongyi Zhou, Prof. Shuwang Lü, Acad. Zhengyao Wei, Acad. Changxiang Shen, Acad. Yongnian Lin, Acad. Binxing Fang, Acad. Guangnan Ni, Acad. Andrew C. Yao, Acad. Jinpeng Huai, Acad. Xicheng Lu, Prof. Jie Wang, Prof. Zhiying Wang, Acad. Wenhua Ding, Acad. Xiangke Liao, Prof. Ping Luo, Rese. Baodong Zhang, Rese. Shizhong Wu, Prof. Yixian Yang, Prof. Maozhi Xu, Prof. Zhiqiu Huang, Prof. Zhenmin Tang, Prof. Jianfeng Ma, Prof. Dingyi Pei, Prof. Mulan Liu, Prof. Lequan Min, Prof. Bogang Lin, Prof. Renji Tao, Prof. Quanyuan Wu , and Prof. Zhichang Qi for their important suggestions, corrections, and helps.